Yu.A.Troyan*, A.V.Beljaev, A.Yu.Troyan, E.B.Plekhanov, A.P.Jerusalimov, G.B.Piskaleva, S.G.Arakelian .
*E-mail:atroyan@jinr.ru

# The Search and Study of the Baryonic Resonances with the Strangeness $S = +1$ in the System of $nK^+$ from the Reaction $np \to npK^+K^-$ at the Momentum of Incident Neutrons $P_n = (5.20 \pm 0.12) GeV/c$

The production and properties of the resonances with the strangeness $S = +1$ in the system of $nK^+$ were studied in the reaction $np \to npK^+K^-$ at the momentum of incident neutrons $P_n = (5.20 \pm 0.12) GeV/c$.

A number of peculiarities were found in the effective mass spectrum of the mentioned above system. All these resonances have a large statistical significance. Their widths are comparable with the mass resolution. The estimation of the spins of resonances was carried out and the rotational band connecting the resonances masses and their spins was constructed.

## INTRODUCTION

D.Diakonov, V.Petrov and M.Polyakov have suggested in the papers [1, 2] the development of the sheme of $SU(3)$-symmetry for states with the strangeness $S=+1$. It was claimed the existence of anti-decuplet $\overline{10}$, which included states consisting of 5 quarks $(uudd\bar{s})$. The dynamics of new resonances was based on the model of chiral soliton. This fact gave the possibility to estimate masses, widths and quantum numbers of expected new effects, to suggest the formula of the rotational band that gave a dependence of the resonance masses on their spins. In the paper [1], $\Theta$-resonance at the mass $M=1.530 GeV/c^2$, width $\Gamma \leq 15 MeV/c^2$ and with quantum numbers $Y=2$, $I=0$, $J^P=1/2^+$ is in the vertex of anti-decuplet.

The hypothesis of the authors [1] is discussed in detail many theoretical works the number of which is close to 50. Detailed review of theoretical works one can find in papers [3, 4, 5] together with the number of critical remarks. The quite different approaches for the problem of these resonances are developed in papers [6, 7]:-integration of quarks into diquarks that is accompanied by a rise of superconducting layers [6];-pure quark picture where can arise isoscalar, isovector and isotensor states consisting of 5quarks (both usual and strange) that strongly extends both the possible quantum numbers, resonances masses and probabilities of their decays ( for example, $\Theta$-resonances may have the quantum numbers $1/2^-, 3/2^-, 5/2^-$ ) [7].

The properties of the particles from anti-decuplet predicted in [1, 2] are the following that allow the direct search of effects. These are both comparatively low masses and accessible for a direct measurement widths. That is why there are a number of experimental works [8] about the discover of a resonance with the mass of ~1.540 $GeV/c^2$ and width of 3÷25 $MeV/c^2$ in $nK^+$ or $pK_s^0$-system.

But no experimental works from the mentioned above did not yet observe neither the rotational band nor more than one resonance state. Also neither spin of resonance nor its parity was not determined. This fact is at first due to a small statistic of experiments, insufficient accuracy and some kind of samplings.

In present work we have attempted to study the characteristics of the observed effects more detailed.

## 1. EXPERIMENT

The study was carried out using the data obtained in an exposure of 1-*m* H$_2$ bubble chamber of LHE (JINR) to a quasimonochromatic neutron beam that was constructed in 1972 due to the acceleration of deuterons by synchrophasotron of LHE. The purpose of the neutron experiment was the study of pentaquarks in $\Delta^{++}\pi^+$-system (described below). Quasimonochromatic neutrons $(\Delta P_n / P_n \approx 2.5\%)$ were generated due to the stripping of accelerated deuterons in 1-*cm* Al target placed inside the vacuum chamber of synchrophasotron. Neutrons were extracted from the accelerator at the angle of 0° to the direction of accelerated deuterons.

A cleaning of the neutron beam from charged particles was provided by the magnetic field of accelerator through which neutrons passed about 12 meters before the exit out of synchrophasotron. The bubble chamber was placed at the distance of 120*m* from Al target. The neutron beam was good collimated and has had an angular spread $\Delta\Omega_n \approx 10^{-7} sterad$. The neutron beam had no admixture neither from charged particles nor γ-quanta. The detailed description of the neutron channel was made in [9].

1-*m* H$_2$ bubble chamber was placed inside the magnetic field of ~1.7T. In results we have had a good accuracy of the momenta of secondary charged particles (δ$P$≈2% for protons and δ$P$≈3% for $K^+$ and $K^-$ from the reaction $np \to npK^+K^-$). The angular accuracy was ≤ 0.5°.

The channels of the reactions were separated by the standard $\chi^2$–method taking into account the corresponding coupling equations [10]. There is only one coupling equation for the parameters of the reaction $np \to npK^+K^-$ (energy conservation law) and the experimental $\chi^2$–distribution must be the same as the theoretical $\chi^2$–distribution with one degree of freedom.

Fig. 1a shows the experimental (histogram) and the theoretical (solid curve) $\chi^2$–distributions for the reaction $np \to npK^+K^-$. One can see a good agreement between them up to $\chi^2 = 1$ and some difference for $\chi^2 > 1$. Therefore we have used only the events with $\chi^2 \leq 1$ for the further analysis. 15% of events with this limitation satisfy two hypothesis: channel $np \to npK^+K^-$ and channel $np \to np\pi^+\pi^-$. In this case $\chi^2$–value of the hypothesis of $K$-mesons ("K") is always less than $\chi^2$–value of the hypothesis of $\pi$-mesons ("$\pi$"). All this events are attributed to the "K" hypothesis. A difference between some test distributions for single hypothesis and double hypothesis events was not observed.

Fig. 1b shows missing mass distribution for the events of $\chi^2 \leq 1$. One can see that the distribution has the maximum at the value equal to the neutron mass with accuracy of 0.1 $MeV/c^2$ and is symmetric about the neutron mass. Later on a small number of events the missing masses out of range pointed by arrows were excluded for more purity of data.

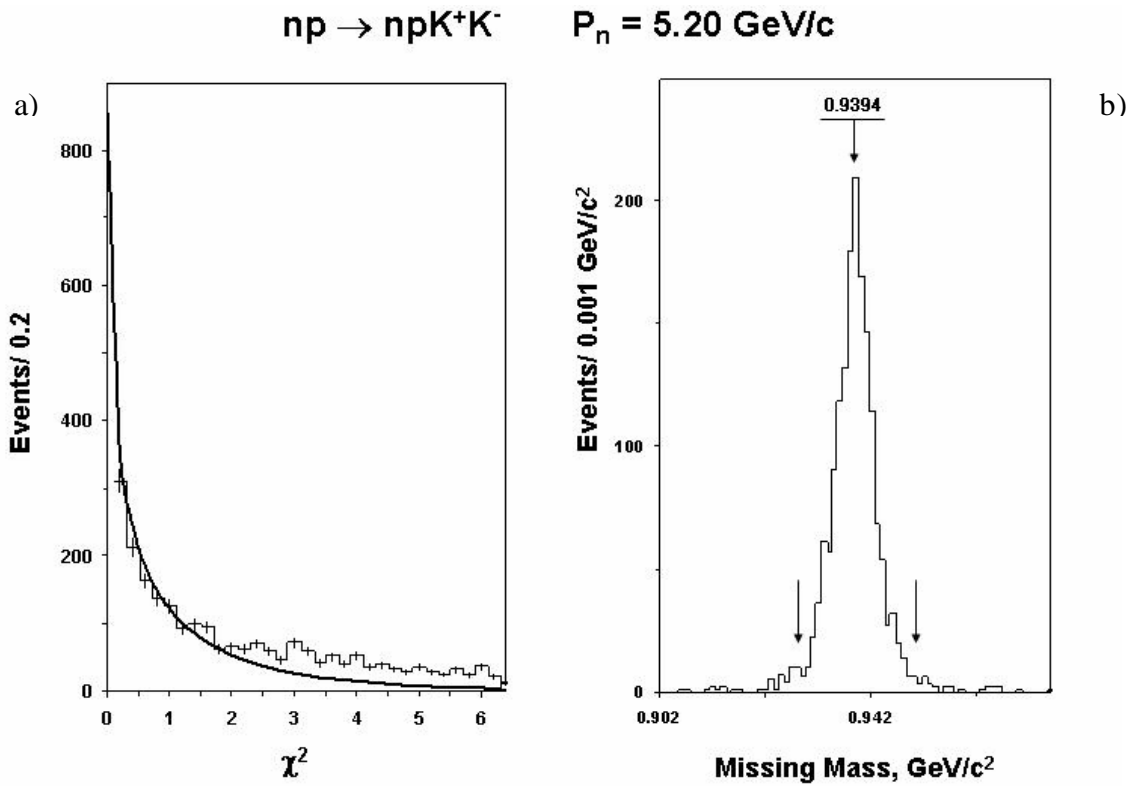

Fig.1
a) the experimental (histogram) and the theoretical (solid curve) $\chi^2$-distribution for the reaction $np \to npK^+K^-$;
b) missing mass distribution for the events of $\chi^2 \leq 1$.

In consequence of this select 1558 events from the reaction $np \to npK^+K^-$ at $P_n = (5.20 \pm 0.12) GeV/c$ under condition of $4\pi$-geometry (the absence of any angular selections).

## 2. RESULTS

Figure 2 shows the effective mass distribution of $nK^+$–combinations for all events from the reaction $np \to npK^+K^-$ at $P_n = (5.20 \pm 0.12)\,GeV/c$. The distribution is approximated by an incoherent sum of the background curve (taken in the form of a superposition of Legendre polynomials up to $8^{th}$ power, inclusive) and by 10 resonance curves taken in the Breight-Wigner form. The part of the background is 75.8 % in this distribution. The requirements to the background curve are the following: firstly, the errors of the coefficients must be not more than 50 % for each term of the polynomial; secondly, the polynomial must describe the experimental distribution after "deletion" of resonance regions with $\overline{\chi^2} = 1.0$ and $\sqrt{D} = 1.4$ (the parameters of $\chi^2$–distribution with 1 degree of freedom). The parameters for the distribution in. fig. 2 are $\overline{\chi^2} = 0.92 \pm 0.29$ and $\sqrt{D} = 1.33 \pm 0.20$. The same parameters for the background curve normalized to 100 % of events (resonance regions are included) are $\overline{\chi^2} = 1.40 \pm 0.19$ and $\sqrt{D} = 2.38 \pm 0.14$. The significance level of the resonance at $M = 1.541\,GeV/c^2$ is 4.5 S.D.

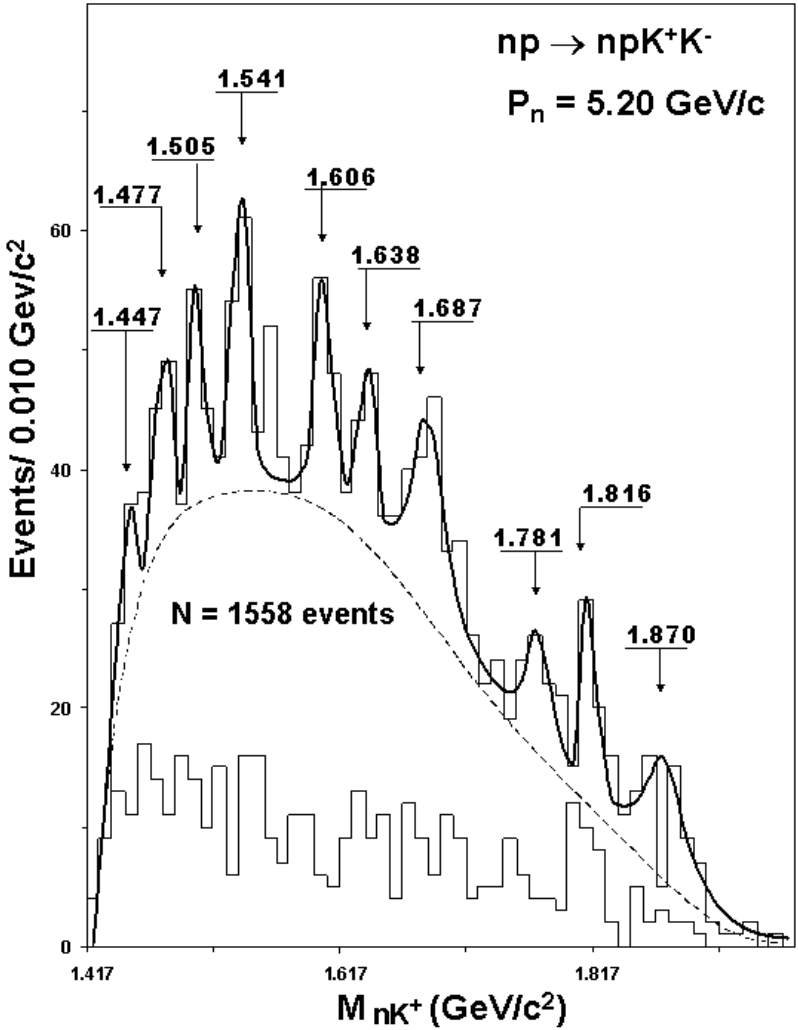

Fig.2
The effective mass distribution of $nK^+$–combinations for all events from the reaction $np \to npK^+K^-$ at $P_n = (5.20 \pm 0.12)GeV/c$.

Dotted line – the background curve taken in the form of Legendre polynomial of $8^{th}$ degree.
Solid line – the sum of the background curve and the 10 resonance curves taken in the Breight-Wigner form.

Lower histogram – the effective mass distribution of $nK^+$–combinations selected under condition $\{\cos\Theta_n^* < -0.85 \cup \cos\Theta_n^* > 0.85\}$.

In the same fig. 2 the distribution of effective masses is presented for $nK^+$–combinations selected under condition $\{\cos\Theta_n^* < -0.85 \cup \cos\Theta_n^* > 0.85\}$, where $\Theta_n^*$-the angle of secondary neutron emission in c.m.s. One can see that this distribution has no essential bumps and a deletion of such kind of events can decrease the background.

Fig. 3 shows the distribution of effective masses of $nK^+$–combinations for the events selected under condition $\{-0.85 < \cos\Theta^*_n < 0.85\}$.

The distribution is approximated by an incoherent sum of the background curve taken in the form of superposition of Legendre polynomials up to $8^{th}$ power and by 8 resonance curves taken in the Breight-Wigner form. The statistical significances somewhat increase for the resonances on the right from the mass $M = 1.541 GeV/c^2$ and somewhat decrease for the narrow resonances on the left one.

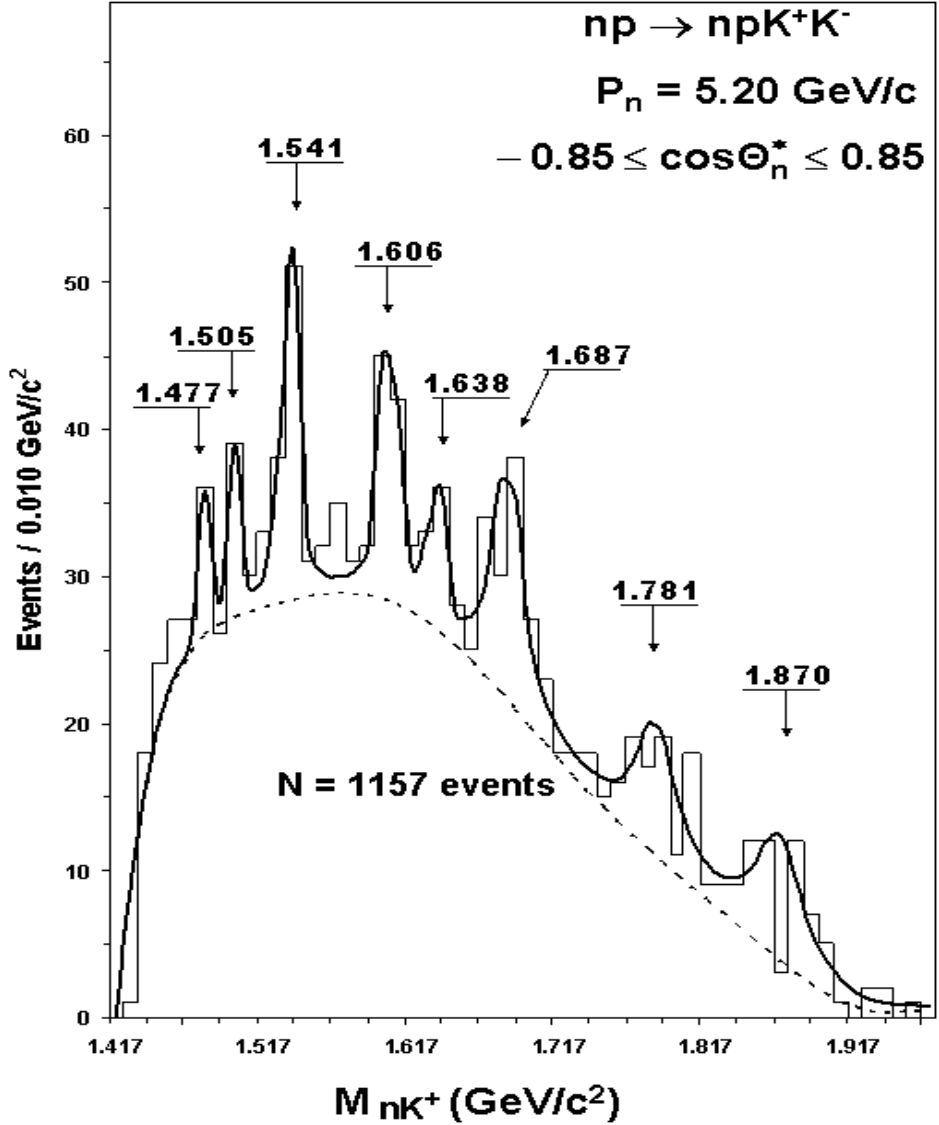

Fig.3

The effective mass distribution of $nK^+$–combinations for the events selected under condition $\{-0.85 < \cos\Theta^*_n < 0.85\}$ from the reaction $np \to npK^+K^-$ at $P_n = (5.20 \pm 0.12) GeV/c$.

Dotted line – the background curve taken in the form of Legendre polynomial of $8^{th}$ degree.

Solid line – the sum of the background curve and the 8 resonance curves taken in the Breight-Wigner form

For a better study of low-masses resonances, the distribution of effective masses of $nK^+$–combinations was constructed with bins of 5 $MeV/c^2$ (up to the mass of ~1.663 $GeV/c^2$). This distribution (Fig. 4) was approximated by an incoherent sum of the background curve taken in the form of a superposition of Legendre polynomials up to $4^{th}$ power and 6 resonance curves taken in the Breight-Wigner form.

The resonance at $M = 1.541 GeV/c^2$ exceed the background by 5.2 S.D , the resonance at the masse of $M = 1.605 GeV/c^2$ – by 5.4 S.D and the resonance at the mass of $M = 1.505 GeV/c^2$ – by 3.1 S.D. The widths of the resonances are more precise determined by means of this distribution (see Tab. I).

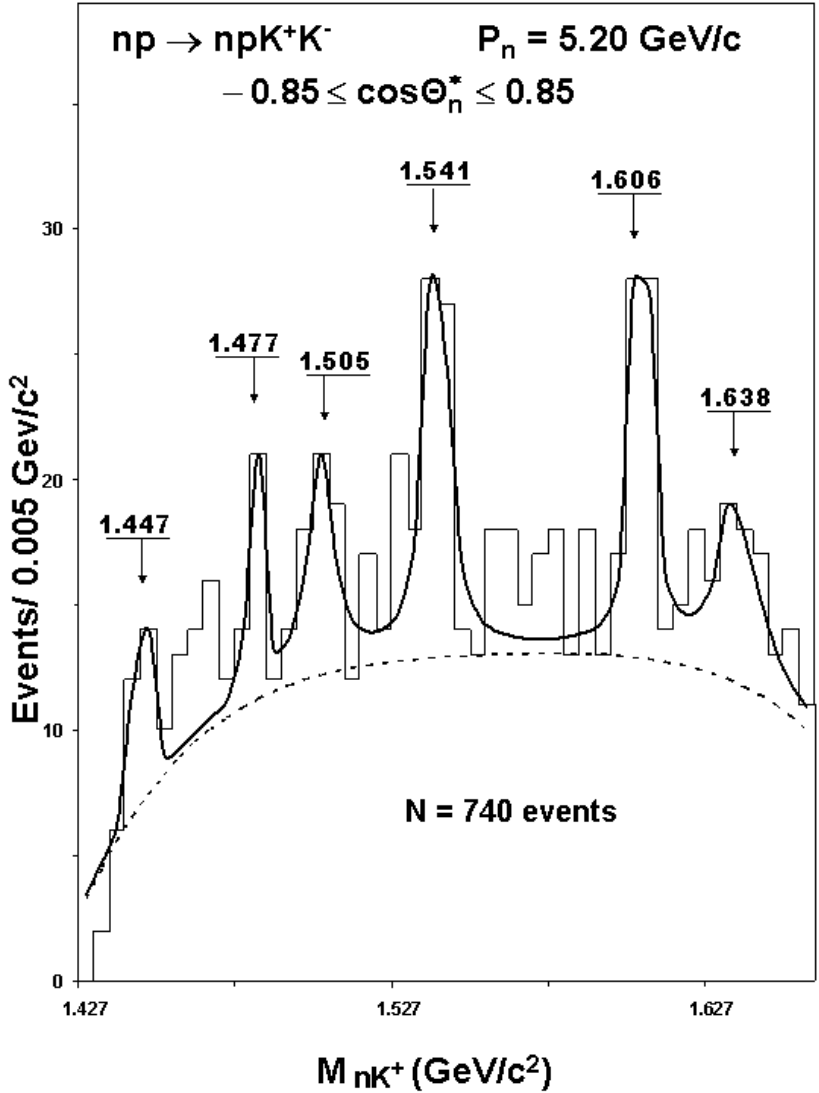

Fig.4
The effective mass distribution of $nK^+$–combinations for the events selected under condition of $\{-0.85 < \cos\Theta_n^* < 0.85\}$ from the reaction $np \to npK^+K^-$ at $P_n = (5.20 \pm 0.12) GeV/c$.

Dotted line – the background curve taken in the form of Legendre polynomial of $4^{th}$ degree.

Solid line – the sum of the background curve and the 6 resonance curves taken in the Breight-Wigner form

We have undertook an attempt to increase the statistical significances of some resonances. This attempt was based under the assumption that resonances were produced by means of $K$-exchange mechanism: a well known resonance ($\Sigma^*$ or $\Lambda^*$) decaying trough $pK^-$-mode was produced in one of the vertices of the corresponding diagram and the resonance in the $nK^+$-system was produced in another vertex. $K^-$-meson from decay of a well known resonance could kinematically be correlated with the resonance in the $nK^+$-system. In consequence kinematically produced peaks could form in the effective mass distribution of $nK^+K^-$-system.

Figure 5 shows the effective mass distribution of $nK^+K^-$-combinations. A number of peculiarities are clearly observed in this distribution. Corresponding resonances decaying through the mode $R \to NK\bar{K}$ are absent in PDG tables. These are just the same kinematic reflections mentioned above. The same fig. 5 shows the effective mass distribution of $nK^+K^-$-combinations constructed under condition that the effective mass $nK^+$-system is within the range of the resonance at $M = 1.541 GeV/c^2$. Two clusters are clear seen in this distribution near the following masses of $nK^+K^-$-system: 2.020÷2.150 $GeV/c^2$ and 2.240÷2.280 $GeV/c^2$. Corresponding clusters exist for resonances in $nK^+$-system at the masses of $M = 1.606 GeV/c^2$ and $M = 1.687 GeV/c^2$.

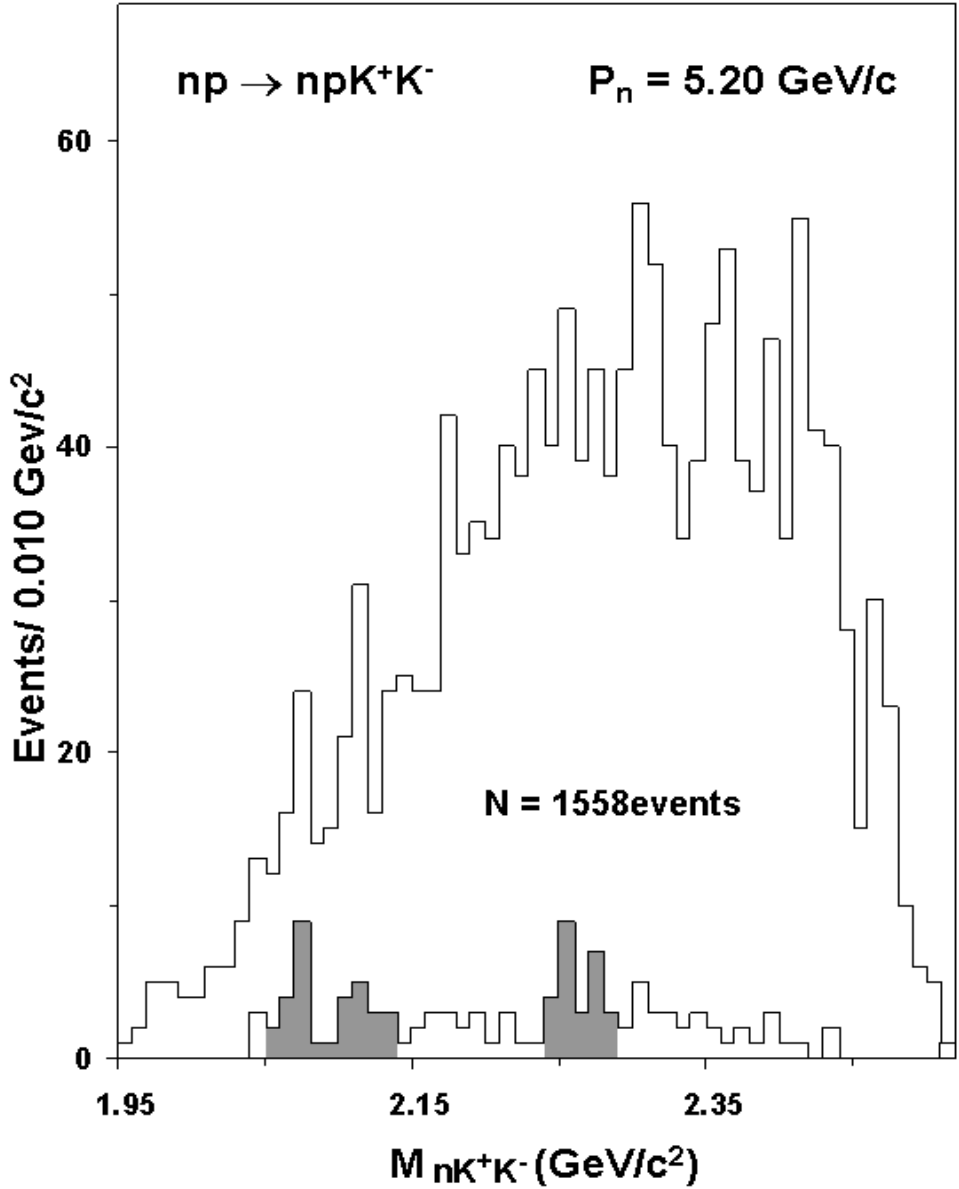

Fig.5

The effective mass distribution of $nK^+K^-$–combinations
from the reaction $np \to npK^+K^-$ at $P_n = (5.20 \pm 0.12) GeV/c$.

Lower histogram – the effective mass distribution of $nK^+K^-$–combinations plotted under condition that the effective mass of $nK^+$–combinations was in the region of the resonance at the mass of $M = 1.541 GeV/c^2$

Selecting regions of masses of $nK^+K^-$-combinations that correspond to the $nK^+$-resonances, we obtain the following distributions of effective masses of $nK^+K^-$-combinations (Fig. 6).

The selected masses of $nK^+K^-$-combinations and the additional limits of the emission angles of secondary neutrons in c.m.s. are shown under each distribution. The additional cut on emission angle somewhat decrease the background but the main effect of enhancement is due to the cut on masses of $nK^+K^-$-combinations.

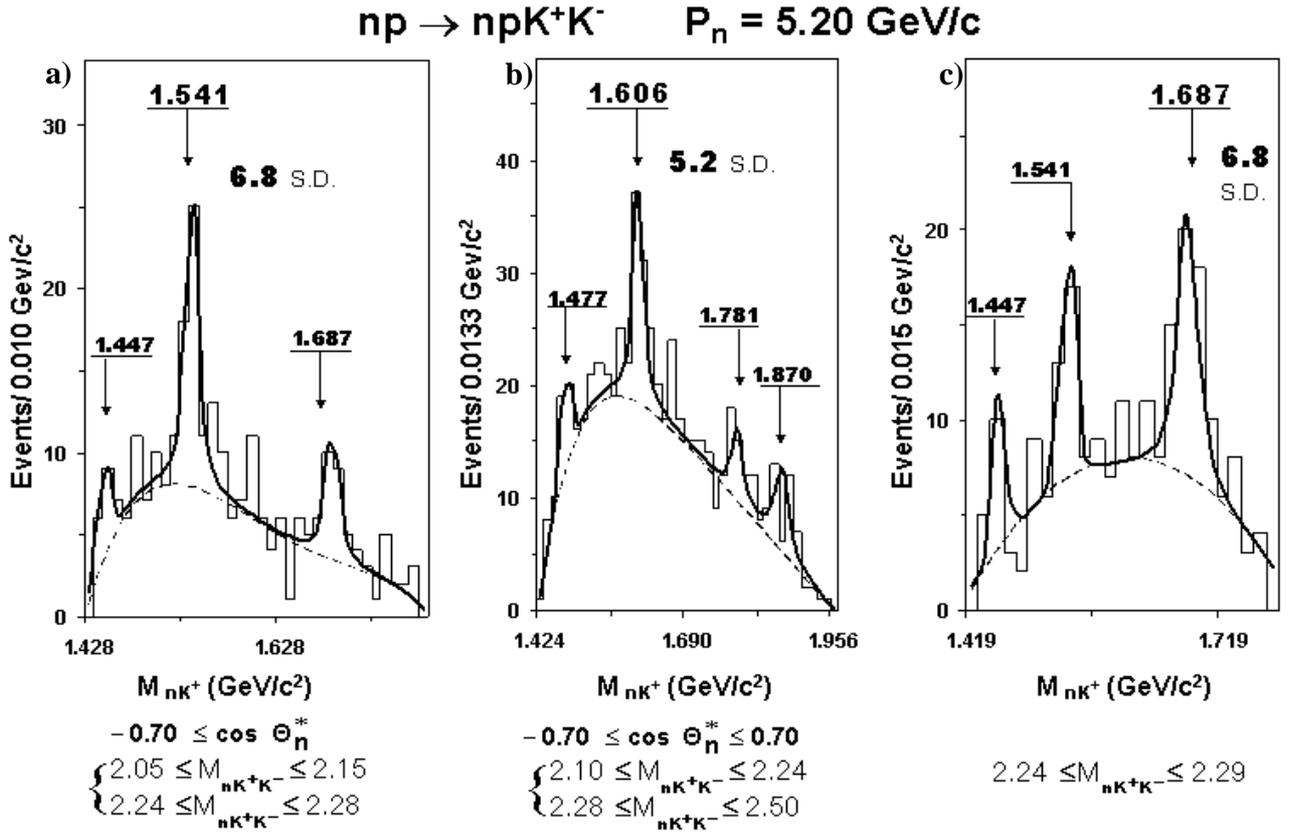

Fig. 6

The effective mass distribution of $nK^+$–combinations from the reaction $np \to npK^+K^-$ at $P_n = (5.20 \pm 0.12) GeV/c$

a) for the resonance at the mass of $M = 1.541 GeV/c^2$;
b) for the resonance at the mass of $M = 1.606 GeV/c^2$,
c) for the resonance at the mass of $M = 1.687 GeV/c^2$.

The selected masses of $nK^+K^-$–combinations and the additional limits of emission angles of secondary neutrons in c.m.s. are shown under each histogram.

Dotted lines – background curves.

Solid lines – approximating curves.

Each of obtained distributions are approximated by an incoherent sum of background curve taken in the form of Legendre polynomial and by resonance curves taken in the Breight-Wigner form.

As a result we get a significant enhancement of effects of three resonances processed by such manner (the values of S.D. are shown in Fig. 6). In this case the number of events in peaks do not decrease as compared with the data presented in Fig. 2, 3 and 4.

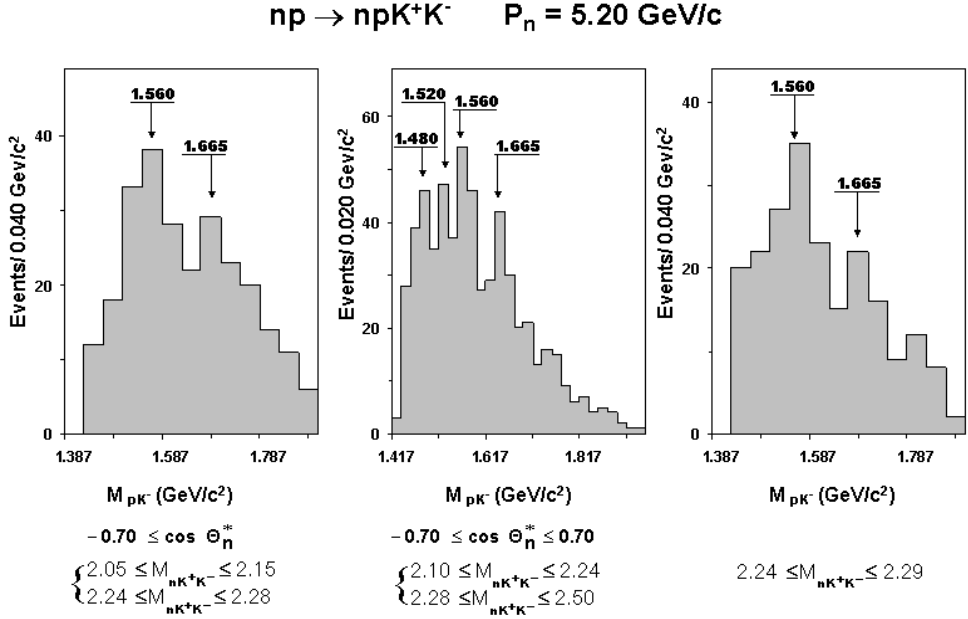

Fig. 7 The effective mass distribution of $pK^-$–combinations from the reaction $np \to npK^+K^-$ at $P_n = (5.20 \pm 0.12) GeV/c$. The selected masses of $nK^+K^-$–combinations and the additional limits of emission angles of secondary neutrons in c.m.s. are shown under each histogram (corresponds to fig. 6).

Figure 7 shows the distributions of effective masses of $pK^-$-combinations under the same conditions of sampling as in Fig. 6. On can observe peculiarities in the masses of $pK^-$-combinations corresponding to well known $\Sigma^*$ or $\Lambda^*$-resonances (these peculiarities are well observed also in the distributions of effective masses of $pK^-$-combinations constructed without limitations mentioned above).

We have tried to estimate the values of spins for the observed resonances in $nK^+$-system. To do this there were constructed the distributions of emission angles of neutron from resonance decays relative to the direction of resonance emission in general c.m.s. All quantities were taken in the rest system of resonance (helicity coordinate system). In the helicity coordinate system, the angular distributions have to be described by the sum of Legendre polynomials of even degrees and a maximum degree has to be equal to $(2J-1)$ where $J$ – spin of a resonance (for half-integer spins). By this manner the value of low limit of resonance spin was estimated. The authors were grateful to Dr. V.L.Luboshits for his help.

Figure 8 shows the angular distributions for six resonances which masses are within the ranges pointed on plots. The backgrounds are constructed using events at the left and at the right of the corresponding resonance band and subtracted using the weight in proportion to a contribution of a background into resonance region. No limitations on emission angles of secondary particles were used by construction of these distributions (cuts on emission angle of secondary neutron do not change the results). It is necessary during the approximation that errors of coefficients Legendre polynomials do not exceed 50%.

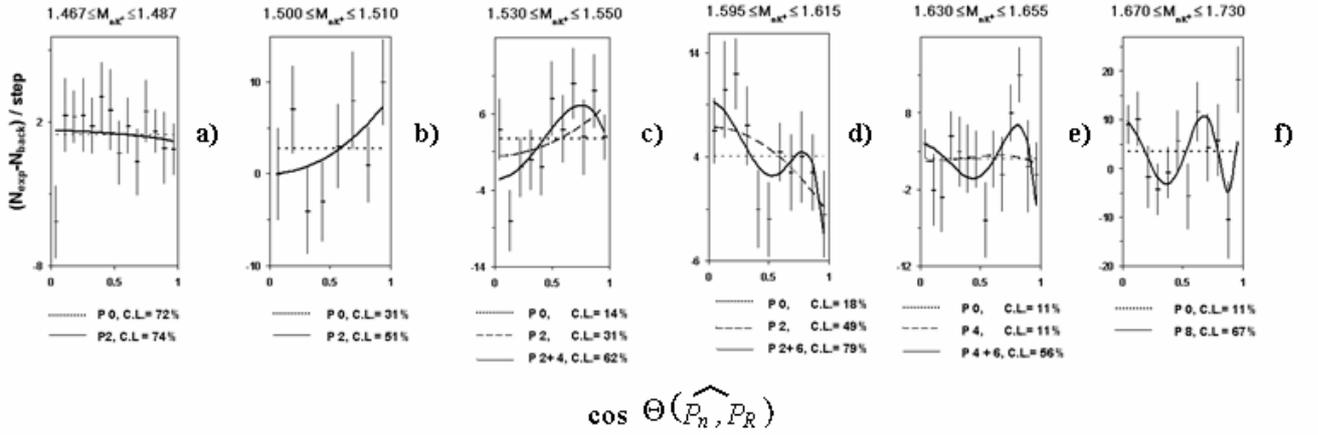

Fig.8 The distribution of the emission angles of secondary neutrons in helicity coordinate system:
  a) for the resonance at the mass of $M = 1.477 GeV/c^2$,
  b) for the resonance at the mass of $M = 1.505 GeV/c^2$,
  c) for the resonance at the mass of $M = 1.541 GeV/c^2$,
  d) for the resonance at the mass of $M = 1.606 GeV/c^2$,
  e) for the resonance at the mass of $M = 1.638 GeV/c^2$,
  f) for the resonance at the mass of $M = 1.687 GeV/c^2$.

One can see in the plots presented in fig. 8 that the distribution for the resonance at $1.467 < M < 1.487 GeV/c^2$ is isotropic and polynomials of high degree are not necessary for an approximation, and therefore its spin is equal to $J \geq 1/2$. The most probable value of the spin is $J \geq 3/2$ for the resonance at the mass of $1.500 < M < 1.510 GeV/c^2$ although the value $J \geq 1/2$ has enough large confidence level. The value of spin $J \geq 1/2$ for the resonance at the mass of $1.530 < M < 1.550 GeV/c^2$ (the most widely discussed in papers) has a confidence level significantly less than the higher one. The most confidence level is for spin of $J \geq 5/2$. The resonance at the mass of $1.595 < M < 1.615 GeV/c^2$ has rather reliable estimation for the value of spin equal to $J \geq 7/2$.

A qualitative estimation can be done studying the shapes of plots: these plots must have $(2J-3)/2$ extrema and a "trivial" one at $\cos\Theta = 0$ that is rather in agreement with the data in fig. 8d.

Figs. 8e, 8f. present results of spin studies for more heavier resonances. The resonance at the mass of $1.630 < M < 1.655 GeV/c^2$ has the value of spin equal to $J \geq 7/2$ with a good confidence level, the resonance at the mass of $1.670 < M < 1.730 GeV/c^2$ has the value of spin equal to $J \geq 9/2$ with a larger confidence level. The high-mass resonance has a weak estimation because of a rather worse mass resolution at this region and a larger influence of the background.

The results are presented in Tab. I.

| $M_{exp} \pm \Delta M_{exp}$ $GeV/c^2$ | $\Gamma_{exp} \pm \Delta\Gamma_{exp}$ $GeV/c^2$ | $\Gamma_R \pm \Delta\Gamma_R$ $GeV/c^2$ | $J_{exp}$ | S.D. |
|---|---|---|---|---|
| $1.447 \pm 0.007$ | $0.005 \pm 0.004$ | $0.004 \pm 0.004$ | | 3.2 |
| $1.467 \pm 0.003$ | $0.008 \pm 0.003$ | $0.008 \pm 0.004$ | | 2.3 |
| $1.477 \pm 0.002$ | $0.005 \pm 0.003$ | $0.002^{+0.006}_{-0.002}$ | 1/2 | 3.0 |
| $1.505 \pm 0.004$ | $0.008 \pm 0.003$ | $0.005 \pm 0.005$ | 3/2 | 3.5 |
| $1.541 \pm 0.004$ | $0.011 \pm 0.003$ | $0.008 \pm 0.004$ | 5/2 | 6.8 |
| $1.606 \pm 0.005$ | $0.014 \pm 0.005$ | $0.011 \pm 0.006$ | 7/2 | 5.2 |
| $1.638 \pm 0.005$ | $0.016 \pm 0.011$ | $0.012^{+0.015}_{-0.012}$ | 7/2 | 3.6 |
| $1.687 \pm 0.007$ | $0.027 \pm 0.007$ | $0.024 \pm 0.008$ | 9/2 | 6.8 |
| $1.781 \pm 0.008$ | $0.029 \pm 0.012$ | $0.023 \pm 0.015$ | | 4.1 |
| $1.870 \pm 0.019$ | $0.036 \pm 0.010$ | $0.032 \pm 0.011$ | | 5.9 |

The first column contains the experimental values of the resonance masses and their errors.
The second column contains the experimental values of the total width of the resonances.
The third column contains the true widths of the resonances and their errors.

The true width of a resonance is obtained by a quadratic subtraction of the value of mass resolution from the experimental value of the width. The function of the mass resolution [11] increase with the increasing of masses:

$$\Gamma_{res}(M) = 4.2\left[\left(M - \sum_{i=1}^{2} m_i\right)/0.1\right] + 2.8 ,$$

where: $M$ – the mass of the resonance, $m_i$ – the masses of the particles composing this resonance.
$M$ and $m_i$ – are in $GeV/c^2$.
Coefficients 4.2 and 2.8 are in $MeV/c^2$.

For example, the value of the mass resolution is equal to $\approx 7$ $MeV/c^2$ for the resonance at the mass of $M = 1.541 GeV/c^2$.

The fourth column contains the values of the spins of resonances or, in more exact terms, the lower limits of spins as it was explained during discussion of the procedure of spins estimation.

Fifth column presents the statistical significances of the resonances determined as the ratio of the number of events in the resonance to the square root of the number of background events under the resonance curve.

The estimation of the cross-section for the resonance at the mass of $M = 1.541 GeV/c^2$ in the $nK^+$-system from the reaction $np \to npK^+K^-$ is $\sigma = (3.5 \pm 0.7) \mu b$ at $P_n = (5.20 \pm 0.12) GeV/c$.

## 3. DISCUSSION

We have tried to systematize the obtained results using the formula for the rotational bands suggested in papers of Diakonov et al. [1, 2]:

$M_J = M_0 + kJ(J+1)$ (1),

where: $M_J$ – the mass of the resonance, $J$ – its spin, $M_0$ – rest mass of the soliton, $k$ – the value equal to the inversed multiplied by twice inertia moment of soliton (we use the terminology of the paper [2])

**Table II**

a)

| $M_0 = 1.462$ GeV/$c^2$ | | $k = 0.0090$ | |
|---|---|---|---|
| $J$ | $M_J$ | $M_{exp} \pm \Delta M_{exp}$ | $J_{exp}$ |
| 1/2 | 1.469 | 1.467 ± 0.003 | |
| 3/2 | 1.496 | 1.505 ± 0.004 | 3/2 |
| 5/2 | 1.541 | 1.541 ± 0.004 | 5/2 |
| 7/2 | 1.604 | 1.606 ± 0.005 | 7/2 |
| 9/2 | 1.685 | 1.687 ± 0.007 | 9/2 |
| 11/2 | 1.784 | 1.781 ± 0.008 | |
| 13/2 | 1.901 | 1.870 ± 0.019 | |

b)

| $M_0 = 1.471$ GeV/$c^2$ | | $k = 0.0107$ | |
|---|---|---|---|
| $J$ | $M_J$ | $M_{exp} \pm \Delta M_{exp}$ | $J_{exp}$ |
| 1/2 | 1.471 | 1.477 ± 0.002 | 1/2 |
| 3/2 | 1.511 | 1.505 ± 0.004 | 3/2 |
| 5/2 | 1.565 | | |
| 7/2 | 1.640 | 1.638 ± 0.005 | 7/2 |
| 9/2 | 1.736 | | |
| 11/2 | 1.854 | 1.870 ± 0.019 | |

When looking to the plots of the effective mass distribution of $nK^+$-combinations one can observe that strong peculiarities are accompanied by more weak one: the weak peculiarity at the mass of $M = 1.467 GeV/c^2$, the bump at the mass region of $M = 1.565 GeV/c^2$ and others. Therefore we have carried out the approximation of the mass distributions versus spin using two variants. Both of them are presented in Tab. II. One can see a good agreement between the experimental data and the formula (1). In Tab. IIa the largest predicted mass at $1.901 GeV/c^2$ $(J = 13/2)$ can be cut by the phase space to the right and be observed experimentally at a lower mass. In Tab. IIb there are absent data about experimental values of masses and spins in third and fifth lines. There are only bumps at these masses that are not provided statistically as resonances.

Taking the inertia moments in the form of $I = M_0 \cdot r^2$ and using the value of $k = 1/2I$ from Tab. II it is possible to determine the radius of soliton. It is equal to $\approx 1,2$ fm that is close to π-meson radius $(\approx 1.35$ fm$)$.

We have done another approximation of the observed rotational bands proposing that the mass of an exited state depends not on of a resonance spin but on its orbital moment $l$:

$M_l = M_0 + kl(l+1)$ (2).

The results are presented in Tab. III:

IIIa – for "strong" resonances and IIIb – for "weak" one. The values of orbital moments are taken arbitrarily but so that they do not contradict the estimations of the spins. Such approximations better satisfy to experimental data. Moreover, the "play" was begun for the resonance at the mass of $M = 1.447 GeV/c^2$ that was observed in the most of our distributions and was discussed in some theoretical studies.

| a) | | |
|---|---|---|
| $M_0 = 1.481$ $GeV/c^2$ | | $k = 0.0100$ |
| $l$ | $M_l$ | $M_{exp} \pm \Delta M_{exp}$ |
| 0 | 1.481 | $1.477 \pm 0.002$ |
| 1 | 1.501 | $1.505 \pm 0.004$ |
| 2 | 1.541 | $1.541 \pm 0.004$ |
| 3 | 1.601 | $1.606 \pm 0.005$ |
| 4 | 1.681 | $1.687 \pm 0.007$ |
| 5 | 1.781 | $1.781 \pm 0.008$ |
| 6 | 1.901 | $1.870 \pm 0.019$ |

| b) | | |
|---|---|---|
| $M_0 = 1.447$ $GeV/c^2$ | | $k = 0.0100$ |
| $l$ | $M_l$ | $M_{exp} \pm \Delta M_{exp}$ |
| 0 | 1.447 | $1.447 \pm 0.007$ |
| 1 | 1.467 | $1.467 \pm 0.003$ |
| 2 | 1.507 | $1.505 \pm 0.004$ |
| 3 | 1.567 | |
| 4 | 1.647 | $1.638 \pm 0.005$ |
| 5 | 1.747 | |
| 6 | 1.867 | $1.870 \pm 0.019$ |

Taking into account the assumption about the orbital moments, the parity of the resonance at the mass of $M = 1.541 GeV/c^2$ is negative. One can conclude taking into account in addition the value of its spin $J = 5/2$ that this resonance is not placed at the vertex of anti-decuplet suggested in papers [1, 2]. But there is a probability that the resonance at the mass of $M = 1.501 GeV/c^2$ (with positive parity and spin equal to 1/2) is placed at the vertex. Our determination of the spin for the mass of $M \approx 1.505 GeV/c^2$ does not contradict to the fact that there can be placed 2 resonances at the mass of $M = 1.501 GeV/c^2$ ($J^P = 1/2^+$) and at the mass of $M = 1.507 GeV/c^2$ ($J^P = 3/2^-$). In this case both of them are very narrow and are shifted relative to each other that gives in result the average value of an experimental mass equal to $M = 1.505 GeV/c^2$. So it is necessary to do very precise on mass resolution and statistics experiments.

It is necessary to do an additional remark.

The problem of pentaquarks arised in early 1960s. Ya.B.Zeldovich and A.D.Saharov [12] were the first to interpret the observed at that time effects in the system of $p\pi^+\pi^+$ as a manifestation pentaquark states. Our first studies [13] concerning this problem have stimulated the realization of the unique neutron beam [9] for 1-m $H_2$ bubble chamber of LHE JINR due to acceleration of deuterons in synchrophasotron of LHE. 1979 there was published our paper [14] about observation of rather narrow $(\Gamma = 43 MeV/c^2)$ resonance in the effective masses of $\Delta^{++}\pi^+$ $(\Delta^-\pi^-)$-combinations at the mass of $M = 1.440 GeV/c^2$ with statistical significance equal to 5.5 S.D. These resonances could be interpreted as five-quark states - $uuu u \bar{d}$ ($ddd d \bar{u}$) for $\Delta^{++}\pi^+$ $(\Delta^-\pi^-)$. In the same paper there was constructed Regge trajectory for states with $J=T$ and was shown that $N$, $\Delta$, $E_{55}$ (observed at the mass of $M = 1.440 GeV/c^2$) were placed on it. The slope of the trajectory was equal to $1.680(GeV/c^2)^{-2}$.

The existence of these new resonances with $J=T$ was predicted in papers of A.Grigorian and A.Kaidalov [15] during the investigation of superconverged sum rules for the reggeon-particle scattering. Their predictions have coincided with our data.

We have published in 1983 the following paper [16] about this problem using the increased statistic. There were additionally observed 2 states at the masses of $M = 1.522 GeV/c^2$ and $M = 1.894 GeV/c^2$. By this means the question about states containing more than 3 quarks is discussed for a long time and there are theories predicting them.

In our opinion the question about the number of quarks is not important in the assumptions of D.Diakonov, V.Petrov and M.Polykov. The symmetry approach does not use the term "quark" at all. This approach is very general and therefore it is more important that the model one.

As far as concerned the experimental situation seems to be very complicate. There is observed only one peak at the mass in the region of 1.530-1.540 $GeV/c^2$ in all experiments where effects in systems of $nK^+$ or $pK_S^0$ were studied. It seems likely to be concerned with a law incident energy, insufficient mass resolution, small statistics and various experimental samples.

It seems for us that most essential is now the observation of at least one additional resonance, the determination of spins of at least two resonances and a precise determination of their widths. The predicted law of an increasing of resonances width due to increase of their spins $\Gamma \sim J^3/M^2$ [2] is very hard. With increasing of spin 5 times a resonance width increase 125 times and it is possible to observe something experimentally only if the masses of resonances will strongly increase that is very hard for an observation and provokes a question about the correctness of a nonrelativistic approach used in the model of chiral soliton.

## Acknowledgements

We are grateful to Dr. V.L. Luboshits for his permanent help in our work during many years.

We are also grateful to Dr. E.A.Strokovsky and Dr. M.V. Tokarev who not only paid our attention to this physical problem but they also permanently provided us with the valid physical information. We are grateful to Dr. E.N.Kladnitskay for some useful remarks.

We are grateful to all research workers who help us in the processing of data: to the laboratory assistants of LHE, to engineers of LIT servicing the required apparatus.

This work was carried out in LHE JINR within the framework of the theme 03-1-0983092/2004.